\documentclass[]{spie}  
\usepackage[]{graphicx}

\title{Low transverse emittance electron bunches from two-color laser-ionization injection} 
\author{Lu-Le Yu\supit{a}\supit{b}, Eric Esarey\supit{a}, Jean-Luc Vay\supit{a}, Carl B. Schroeder\supit{a}, Carlo Benedetti\supit{a}, Cameron G. R. Geddes\supit{a}, Sergey G. Rykovanov\supit{a}, Stepan S. Bulanov\supit{b}, Min Chen\supit{c} and Wim P. Leemans\supit{a}\supit{b}
\skiplinehalf
\supit{a}Lawrence Berkeley National Laboratory, Berkeley, California 94720, USA; \\
\supit{b}University of California, Berkeley, California 94720, USA;
\\
\supit{c}Shanghai Jiao Tong Univeristy,  Shanghai 200240, China
}

\authorinfo{Further author information: (Send correspondence to Eric Esarey and Lu-Le Yu)\\Eric Esarey: E-mail: EHEsarey@lbl.gov, Telephone: 1 510 486 5925\\ 
Lu-Le Yu: E-mail: LuleYu@lbl.gov, Telephone: 1 510 495 2267}

\begin{document} 
  \maketitle 

\begin{abstract}
A method is proposed to generate low emittance electron bunches from two color 
laser pulses in a laser-plasma accelerator. A two-region gas
structure is used, containing a short region of a high-Z gas (e.g., krypton) for 
ionization injection, followed by a longer region of a low-Z gas for post-acceleration.
A long-laser-wavelength (e.g., 5 $\mu$m) pump pulse excites plasma wake
without triggering the inner-shell electron ionization of the high-Z gas due to low electric fields. A short-laser-wavelength (e.g., 0.4 $\mu$m) injection pulse,  
located at a trapping phase of the wake, ionizes the inner-shell electrons of the high-Z gas,
resulting in ionization-induced trapping. Compared with a single-pulse ionization injection,
this scheme offers an order of magnitude smaller residual transverse momentum of the electron bunch, which is a result of the smaller vector potential amplitude of the injection pulse. 
\end{abstract}


\section{INTRODUCTION}
\label{sec:intro}  
In laser-driven plasma-based accelerators (LPAs)\cite{Esarey09}, the accelerating and focusing fields (wakefields) are driven  by the ponderomotive force of the laser pulse $F\sim\nabla a^2$, where $a^2=7.3\times10^{-19}[\lambda(\mu m)]^2 I$(Wcm$^{-2}$) (linear polarization) is the normalized laser intensity, $a=eA/m_ec^2$ is the normalized amplitude of the laser vector potential, and $\lambda$ is the laser wavelength in vacuum.
The accelerating field is on the order of
$E$(V/m)$\simeq96\sqrt{n_0(\mathrm{cm^{-3}})}$ with $n_0$ the plasma electron density, which can be several orders of magnitude greater
than those in conventional accelerators. In addition, LPAs have the potential to produce extremely short electron bunches with durations $\tau_b<\lambda_p/c$, where $\lambda_p(\mu m)\simeq3.3\times10^{10}/\sqrt{n_0(\mathrm{cm^{-3}})}$ is the plasma wavelength. In 2004,
high-quality electron bunches with energy $\sim 100$~MeV were produced \cite{Mangles04, Geddes04, Faure04} with significant charge ($>$100 pC), small energy spread ($<$10\%), and low divergence (few mrad). In 2006, high-quality GeV-class electron beams were first demonstrated by using a capillary plasma-channel guided laser in cm-scale plasma\cite{Leemans06b}. Controlled injection methods, such as colliding pulses injection
\cite{Esarey97a,Schroeder99b,Fubiani04,Kotaki04,Faure06,Rechatin09,Kotaki09},
density transitions\cite{Geddes08,Schmid10,Faure10}, and ionization
injection\cite{MinChen06, McGuffey10, Pak10, Liu11, Pollock11, 
MinChen12} are actively being pursued to improve the quality and stability of the electron beams. Experiments have so far successfully demonstrated production of electron bunches with energy spread at $\sim$1\%\cite{Rechatin09,Wiggins2010}, and estimated transverse emittance at $\sim$ 0.1 mm mrad\cite{Plateau2012}. Such high-quality electron beams could be good candidates to drive free-electron lasers (FELs)\cite{Fuchs2009}, which provide a new generation of low-cost, compact light sources\cite{Huang2012,Maier2012}.  Further decreasing the transverse emittance is critical to many applications of the high-energy electron beams, such as drivers future x-ray FELs.

The normalized transverse emittance can be estimated as $\epsilon_n\simeq\sigma_x\sigma_{p_x}/(m_ec)$, where $\sigma_x$ is the root-mean-square (rms) bunch radius and $\sigma_{p_x}/(mc)$ is the normalized rms transverse momentum. In the self-trapping bubble regime, simulations show  $\sigma_x$$\sim$ 0.1 $\mu$m and $\sigma_{p_x}/(m_ec)$$\sim$ 1, resulting in $\epsilon_n$$\sim$ 0.1 mm mrad\cite{Plateau2012}. Recently, simulations show that an electron bunch with excellent normalized emittance (0.04 mm mrad) can be produced via laser ionization in a beam-driven plasma bubble regime\cite{Hidding2012}. In this scheme, a mixture of Li and He gas is used. An electron beam is propagating in the gas mixture with fully-ionized Li, generating an intense Li plasma bubble. The electric field of the electron beam driver is choosen not high enough to ionize the He gas. A tightly-focused (laser spot size $w_0=4$ $\mu$m) nonrelativistic-intensity ($a_0$=0.018)  laser pulse is used to ionize He component and release electrons directly into the accelerating and focusing phase of the Li plasma bubble. Assuming the produced electron rms bunch radius is $\sim w_0/\sqrt{2}$ and the normalized rms transverse momentum is $\sim a_0/2$, the expected minimal emittance is $\sim w_0a_0/2^{3/2}$. In contrast, conventional methods for ionization injection in LPAs, where $a_0\sim2$ and $w_0\sim10$ $\mu m$ are needed to drive the wake, typically produce emittances of a few mm mrad.

\begin{figure}
\begin{center}
\begin{tabular}{c}
\includegraphics[height=7cm]{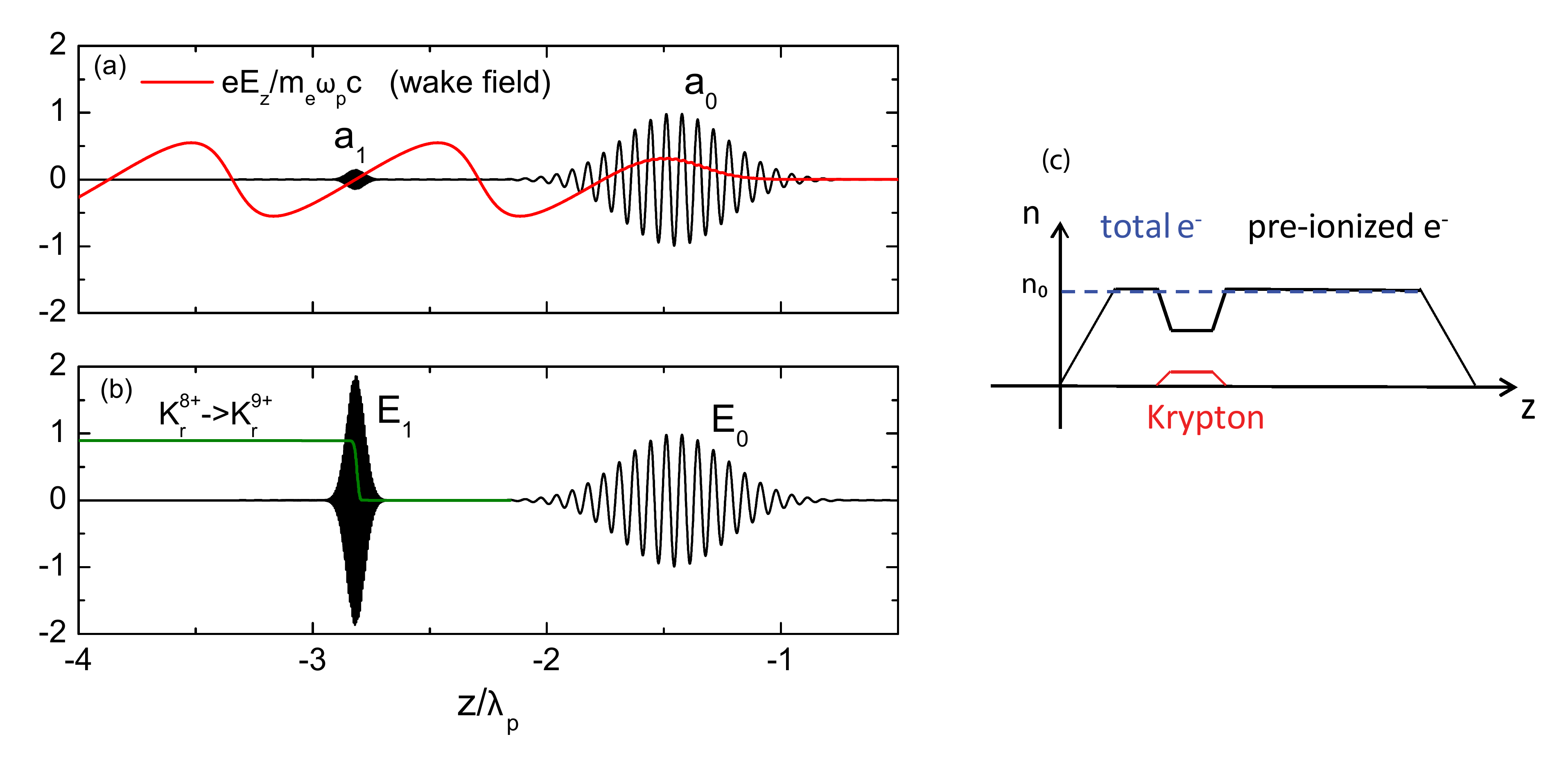}
\end{tabular}
\end{center}
\caption[example] {(Color online) Concept of two-color laser-ionization injection. The high-Z gas is krypton. (a) The vector potentials of the pump pulse $a_0$ (with $\lambda_0=5~\mu$m) and injection pulse $a_1$ (with $\lambda_1=0.4~\mu$m) (black curves) and the normalized excited wakefield $eE_z/m_e\omega_p c$ (red curve). (b) The electric fields of the pump pulse $E_0$ and injection pulse $E_1$ (black curves) and the ionization degree of $K^{8+}_r\rightarrow K^{9+}_r$ (green curve). (c) Schematic profile of the longitudinal density. Electron density  $n_0(x)=n_e(x)+8n_{K_r}(x)$=constant.}
\label{Fig1}
\end{figure}

Here we propose a method to generate low-emittance high-quality electron bunches in a laser-plasma accelerator. As shown in Fig.~\ref{Fig1}(a) and (b), a long-wavelength (e.g., 5 $\mu$m) pump pulse with an amplitude $a_0\simeq1$ propagates in a high-Z gas (e.g., krypton) and ionizes the gas to a mid-charge state, exciting an intense wakefield. The electric field of the pump pulse $E_0$ is relatively low due to the long laser wavelength $a=eE\lambda/(2\pi m_ec^2)$, so that it can not ionize the high-ionization-threshold electrons. A short-wavelength (e.g., 0.4 $\mu$m) injection pulse with an  amplitude $a_1\sim 0.15$ is used to ionize the high-ionization-threshold electron (e.g., $K^{8+}_r\rightarrow K^{9+}_r$) because of the higher electric field $E_1>E_0$. The electrons ionized by the injection pulse at proper wake phases can be trapped and be accelerated to high energy. The gas structure used in the Particle-In-Cell (PIC) simulations in this paper is shown in Fig.~\ref{Fig1}(c). A high-Z gas (krypton) is used for the ionization injection and a low-Z gas is used for the post-acceleration without additional trapping. This two-region gas structure allows us to control the injection number and beam quality of the electron bunch by changing the gas composition, concentration, and length of the high Z gas region\cite{MinChen12}. The total ionized electron number (until the 8th electron of krypton) is fixed and the electron density is set to be $n_0(x)=n_e(x)+8n_{K_r}(x)$=constant, so that the density ramp effects are negligible. The normalized rms transverse momentum of the injected electrons in this scheme is $\sim0.03$, which is an order of magnitude smaller than that in a single pulse ionization injection\cite{MinChen12}.

This paper is organized as follows. Section~\ref{sec:ionization and trapping} presents the high-Z gas selection and the trapping condition for the ionized electron by using a Hamiltonian approach. Section~\ref{sec:pic} presents one-dimensional (1D) PIC simulations of the ionization injection. Section~\ref{sec:summary} gives the conclusions and a discussion of multi-dimensional effects.

\section{High-Z gas selection and electron trapping condition} 
\label{sec:ionization and trapping}
In a single laser pulse ionization injection, typically the peak of the laser pulse is at $a_0\sim2$ for a resonant Gaussian laser pulse and uses an appropriate high-Z gas (e.g., nitrogen), with a laser wavelength of 0.8 $\mu$m\cite{MinChen12}. In this case, ionization injection requires two conditions: (1) the laser intensity needs to be intense enough to excite a sufficiently large wakefield so that an electron ionized at rest near the peak intensity of the laser pulse will be on a trapped orbit; and (2) the ionization threshold of the inner shell electron of the high-Z gas  needs to be close to the laser pulse intensity peak.

In the two-color laser-ionization injection scheme, $a_1$ can be delayed to a position in the wake with a lower trapping threshold. Once the electron is born at rest in the wake phase where the separatrix is negative, it can be trapped. In order to achieve a small transverse emittance of the eletron bunch, the amplitude $a_1$ needs to be as small as possible. Thus, an appropriate gas needs to be chosen so that the pump pulse will not trigger the ionization injection but the injection pulse can generate electrons on trapped orbits with a small $a_1$ due to the short laser wavelength.  

\subsection{Electron Dynamics and Trapping Condition}\label{subsec:trapping condition}
The nonlinear plasma wave generated by the intense circularly-polarized pump pulse in cold
underdense plasma in 1D limit can be written as \cite{Esarey09}
\begin{equation}\label{eq1}
k^2_p\frac{\partial^2\phi}{\partial \xi^2}=
\gamma^2_p\left\{\beta_p\left[1-\frac{1+a_0^2}{\gamma^2_p(1+\phi)^2}\right]^{-1/2}-1\right\}.
\end{equation}
Here $\xi=z-v_pt$ is forward co-moving coordinate,
$\phi(\xi)=e\Phi/mc^2$, $k_p=2\pi/\lambda_p$ and $\beta_p=v_p/c$ are
the normalized potential, the wave number and the normalized phase
velocity of the plasma wave, respectively.  In underdense plasma
$\beta_p\simeq \beta_g\simeq 1$ with $\beta_g$ the group velocity of
the laser pulse.  $\gamma_p=(1-\beta^2_p)^{-1/2}$, and $a_0(\xi)$ is the
pump pulse profile.  The normalized wakefield is
$E_ze/(m_e\omega_pc)=-k_p^{-1}\partial\phi/\partial\xi$. In the absence of the pump pulse, the electron motion in the plasma wake 
can be described using a Hamiltonian approach\cite{Schroeder99b}
\begin{equation}\label{eq2}
H(u_z,\psi)=(1+u^2_z)^{1/2}-\beta_pu_z-\phi(\psi),
\end{equation}
where $\psi=k_p\xi$ is the wake phase and $u_z=p_z/m_ec$ is the normalized longitudinal momentum of the electron.
The separatrix orbit between trapped and untrapped orbits is given by $u_s=u_z(H_s(\gamma_p,\psi_{min}))$,
where $\phi(\psi_{min})=\phi_{min}$ and $H_s=1/\gamma_p-\phi_{min}$.
For the electron ionized inside the injection pulse, the Hamiltonian can be written as\cite{MinChen12}
\begin{equation}\label{eq5}
H_i=1-\phi(\psi_i),
\end{equation}
assuming that the electron is born at rest and $a_1^2\ll1$. Here $\psi_i$ is the wake phase. The wake induced by the injection pulse is neglected since it is typically much smaller than $\phi$. Once the ionized electron lies above the wake separatrix, it can be trapped. Thus, the trapping condition for the ionized electron is
\begin{equation}\label{eq6}
H_i\leq H_s.
\end{equation}

\begin{figure}
\begin{center}
\begin{tabular}{c}
\includegraphics[height=7cm]{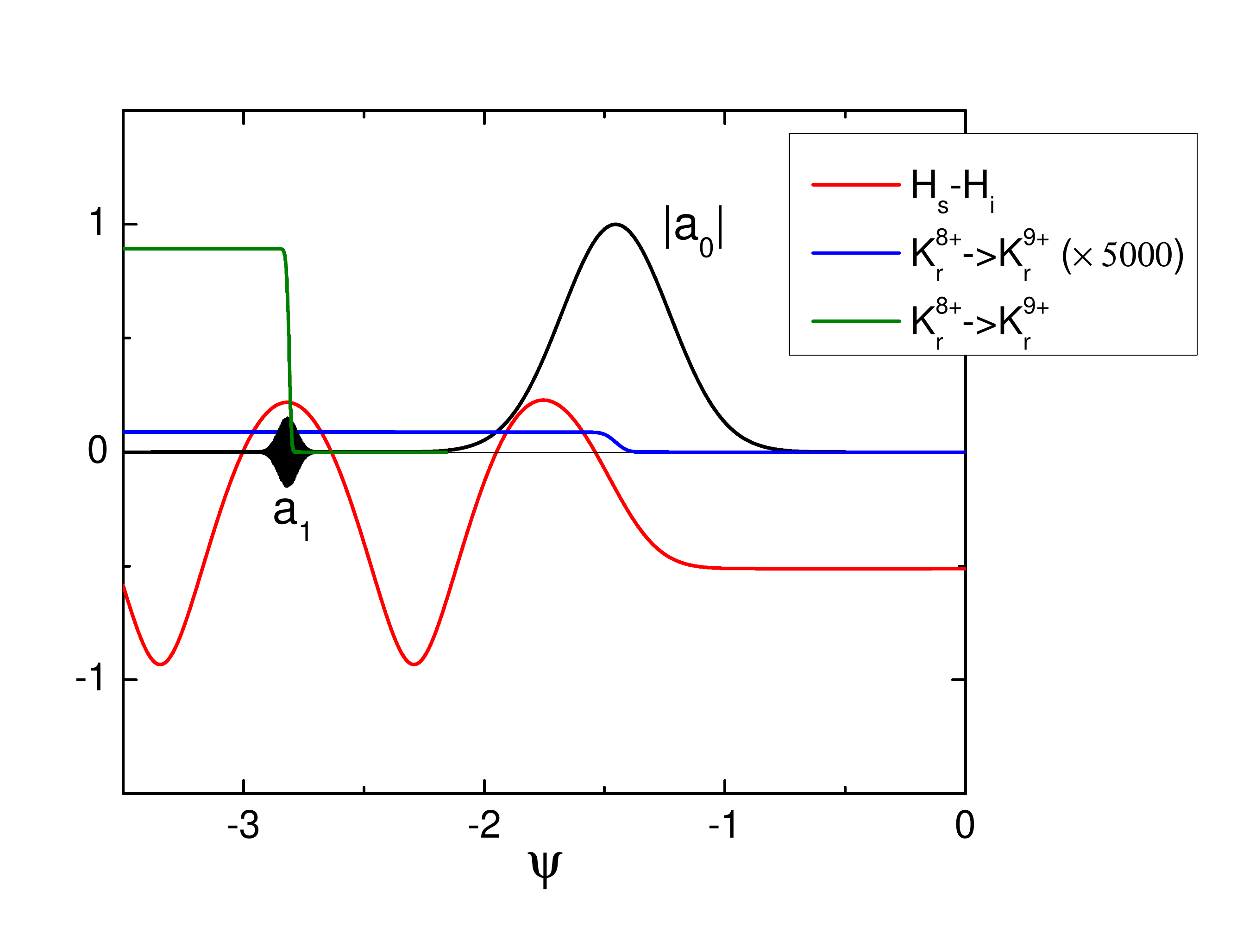}
\end{tabular}
\end{center}
\caption[example] {(Color online) Trapping condition for an ionized electron is $H_s-H_i>0$. $H_s$ and $H_i$ are the Hamiltonian values of the wake separatrix and the ionized electron, and $H_s-H_i$ is shown as a function of wake phase $\psi$ (red curve). Ionization degree of $K^{8+}_r\rightarrow K^{9+}_r$ for the pump pulse with a circular polarization (blue curve) and for the injection pulse with a linear polarization (green curve). The laser pulses have Gaussian profiles $a(\xi)_{0,1}=a_{0,1}\exp[-(\xi-3L_{0,1})^2/L^2_{0,1}]$, with $a_0=1$ and $a_1=0.15$. The laser wavelengths are $\lambda_0=5~\mu$m and $\lambda_1=0.4~\mu$m, and the pulse durations (FWHM) are $\tau_0=1.1774L_0/c=92$ fs and $\tau_1=1.1774L_1/c=16$ fs. The electron density is $n_0=2\times10^{17}$~cm$^{-3}$.}
\label{Fig2}
\end{figure}

By numerically solving Eq.~(\ref{eq1})-(\ref{eq6}), we obtain
the Hamiltonian of the wake separatrix and the ionized electron, as shown in Fig.~\ref{Fig2}.  
Here the initial uniform electron density is $n_0=2\times10^{17}$~cm$^{-3}$, and the pump laser pulse is circularly polarized with a Gaussian profile $a_0(\xi)=a_0\exp[-(\xi-3L_0)^2/L^2_0]$ with $a_0=1$. The laser wavelength of the pump pulse is $\lambda_0=5~\mu$m. The length of the pump pulse is matched with the plasma density to maximize the amplitude of the wakefield by setting $k_pL_0=2$ and the FWHM duration of the pump pulse is $\tau_0=92$ fs. As shown in Fig.~\ref{Fig2}, the initial ionization phase for trapping is where $H_s-H_i>0$, and the optimal injection phase is where $H_s-H_i$ is maximum. The intensity threshold of the pump pulse for trapping at this phase is $a_{0,th}=0.88$. The green curve shows the ionization degree of $K^{8+}_r\rightarrow K^{9+}_r$ for the injection pulse. Here the injection pulse is linearly polarized and has the same profile as the pump pulse. The amplitude, wavelength and duration (FWHM) of the injection pulse are $a_1=0.15$, $\lambda_1=0.4~\mu$m and $\tau_1=16$ fs, respectively.  
It is found that the ionization phase of the injection pulse meets the trapping condition and all the ionized electrons will be trapped in the wake. However, the ionization phase of the pump pulse is where $H_s-H_i<0$ (blue curve), which does not satisfy the trapping condition. Thus, there is no ionization injection from the pump pulse. 

The normalized transverse momentum of the ionized electron is
$u_{\bot}(\psi)=p_{\bot}/m_ec=a_{1,\bot}(\psi)-a_{1,\bot}(\psi_i)$.
Most of the electrons are born at the peak electric field of the laser pulse where $a_{1,\bot}(\psi_i)\cong0$, while some of the electrons are born off-peak with
a finite $a_{1,\bot}(\psi_i)$. Assuming the electron is trapped behind the injection laser pulse, i.e., $a_{1,\bot}(\psi)=0$, then the transverse momentum
is $u_{\bot}(\psi)=a_{1,\bot}(\psi_i)$. This is the residual transverse momentum that results from the electron being born off-peak of the laser pulse, which contributes the initial transverse emittance of the electron bunch. 

\subsection{High-Z Gas Selection}
\begin{figure}
\begin{center}
\begin{tabular}{c}
\includegraphics[height=7cm]{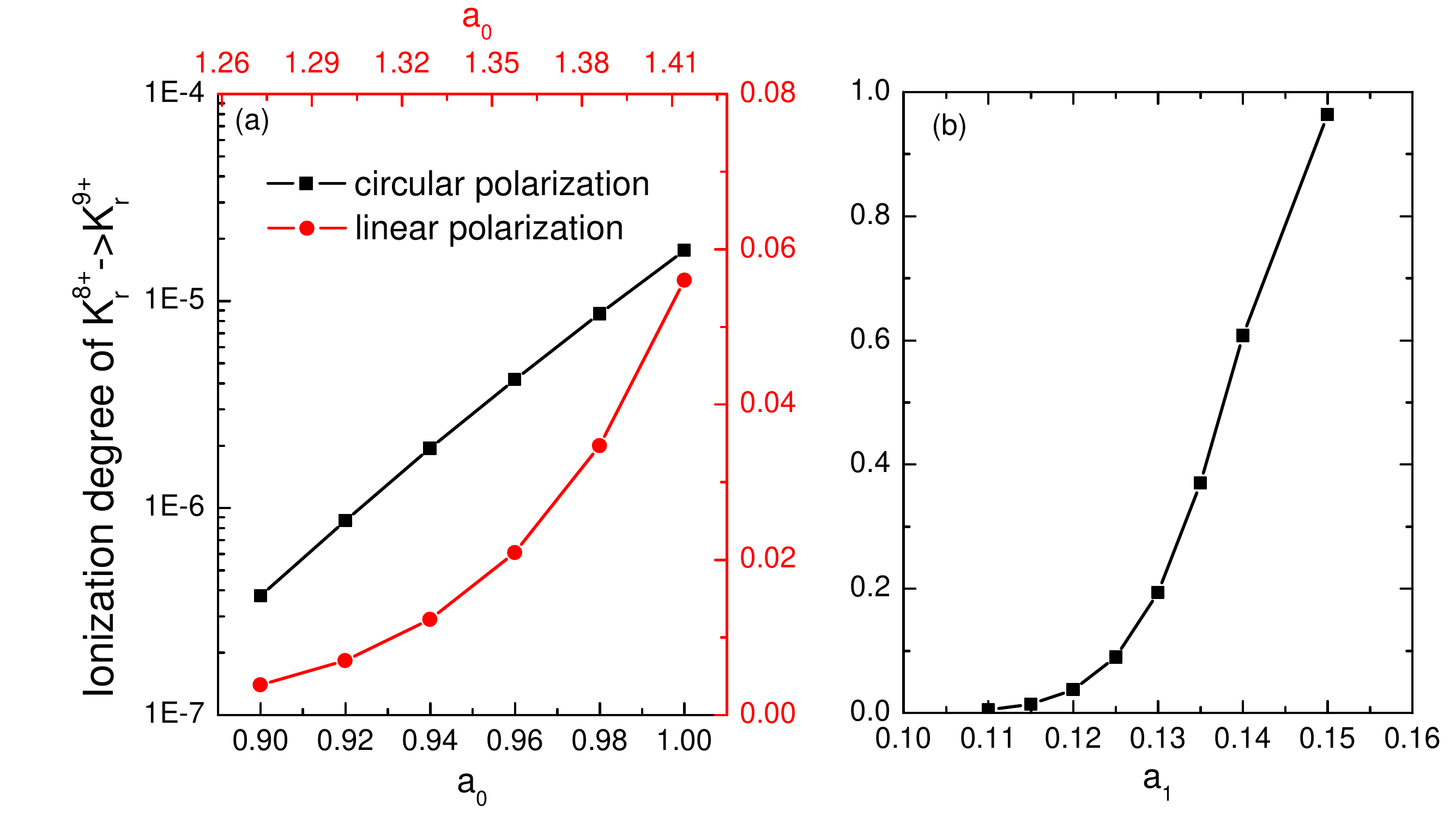}
\end{tabular}
\end{center}
\caption[example] {(Color online) Ionization degree of $K^{8+}_r\rightarrow K^{9+}_r$  versus (a) vector potential amplitude and polarization of the pump pulse [circular polarization (black axes), linear polarization (red axes)] and (b) vector potential amplitude of the injection pulse with linear polarization. Other laser parameters are the same as Fig.~\ref{Fig2}.}
\label{Fig3}
\end{figure}

By using direct current (DC) tunneling  ionization model\cite{Popov2004,MinChen2013}, we found that kyrpton gas may be used for the two- color laser-ionization injection scheme with the pump laser wavelength $\lambda_0=5$ $\mu$m and the injection laser wavelength $\lambda_1=0.4$ $\mu$m. The laser parameters are the same as presented in Fig.~\ref{Fig2}. The ionization potential (IP) of the 8th (IP 126 eV) electron of kyrpton that produces $K_r^{8+}$ is quite low, so that both the pump pulse and injection pulse can fully ionize the gas up to this shell. However, for the 9th (IP 230 eV) electron of Kyrpton that produces $K_r^{9+}$, the ionization degree is strongly dependent on the laser vector potential and laser wavelength (i.e., the laser electric field) for the parameters under consideration. As shown in Fig.~\ref{Fig3}(a), the ionization degrees of $K^{8+}_r\rightarrow K^{9+}_r$ for a circularly-polarized pump pulse is $\geq3$ orders of magnitude smaller than those for a linearly-polarized pump pulse, while keeping the same wake generation in both cases. This is because in order to generate a same wake, the amplitude of a linearly-polarized pulse $a_0$ needs to be $\sqrt{2}$ larger compared to a circularly-polarized pulse, resulting in a much higher ionization degree. Also, the electron motion in a linearly-polarized intense laser pulse with $a_0>1$ is more nonlinear than the one in a circularly-polarized laser pulse with $a_0\leq 1$. Thus, a circularly-polarized pump pulse is choosen in order to avoid the ionization injection in the first bucket of the wakefield. Figure~\ref{Fig3}(b) shows the ionization degrees of $K^{8+}_r\rightarrow K^{9+}_r$ for a linearly-polarized injection pulse. It is found that high ionization degree can be achieved with a small $a_1$ due to the short laser wavelength. For example, the ionization degree is 96\% for $a_1=0.15$. With the decrease of $a_1$, the ionization degree decreases and the ionized electron number decreases, resulting in a smaller injection number.  

\section{Particle-In-Cell simulations}\label{sec:pic}
\begin{figure}
\begin{center}
\begin{tabular}{c}
\includegraphics[height=6.5cm]{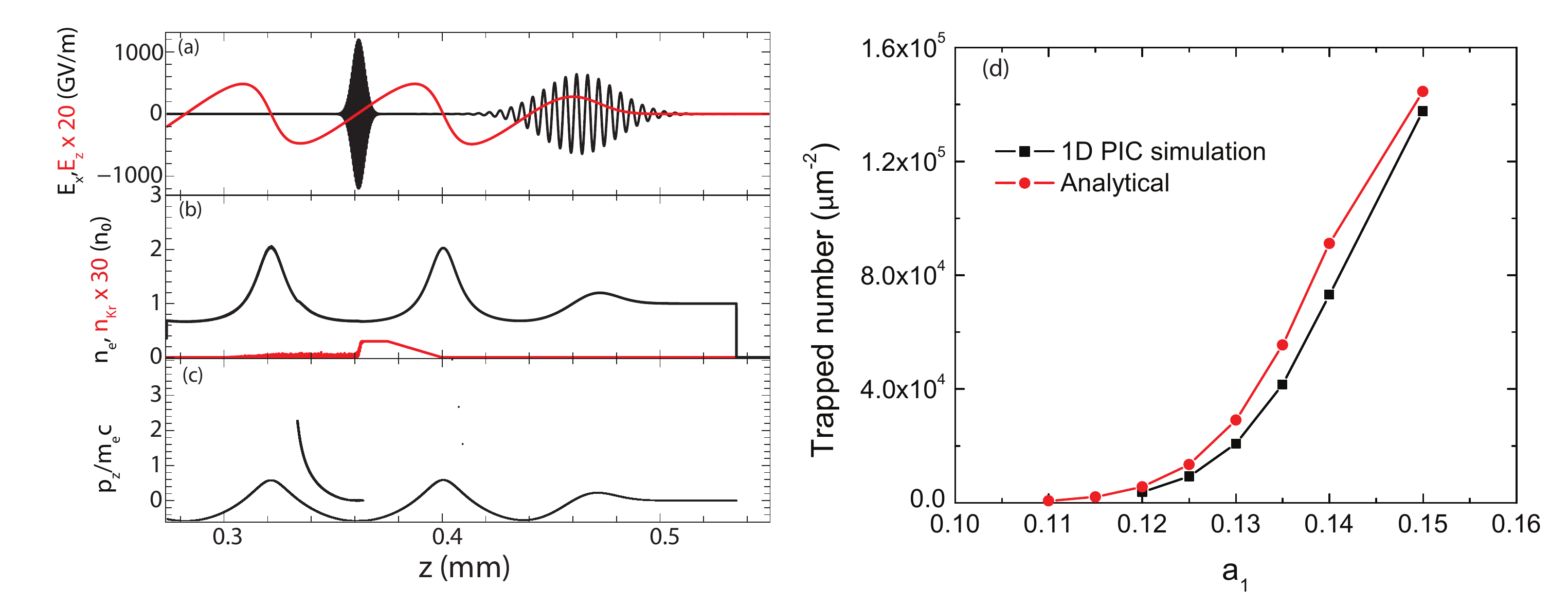}
\end{tabular}
\end{center}
\caption[example] {(Color online)  (a) The electric fields of the two laser pulses (black curves) and the wakefield excited by the pump pulse (red curve). (b) The electron density (black curve) and the krypton density (red curve). (c) The normalized longitudinal momenta of the electrons. (d) The trapped electron number versus the vector potential amplitude of the injection pulse. The electron density and the laser parameters are the same as Fig.~\ref{Fig2}. The krypton concentration is 1\%.}
\label{Fig4}
\end{figure}

The two-color laser-ionization injection was examined via 1D PIC simulations using the WARP code\cite{JeanLuc2012}. In order to save computational time, a mix of pre-ionized electrons and neutral krypton gas is used, as presented in Fig.~\ref{Fig1}(c). The initial electron density $n_0$ is fixed to $n_0=n_e+8n_{K_r}=2\times10^{17}$ cm$^{-3}$ (a plasma wavelength of 75 $\mu$m). The concentration of the krypton in the simulations is set to $n_{K_r}/n_0=1\%$. The length of the trapezoid-shaped krypton gas is 100 $\mu$m with a plateau of 50 $\mu$m. The profiles and durations of the two lasers are the same as described in Fig.~\ref{Fig2}. As discussed above, since the ionization degree of $K_r^{8+}\rightarrow K_r^{9+}$ is much smaller for the circularly-polarized pump pulse, we use a circularly-polarized pump pulse with $a_0=1$ in the simulations. The peak of the pump pulse is located at $z=0$. Figures~\ref{Fig4}(a)-(c) shows the ionization injection process. Here the peak of the injection pulse with $a_1=0.15$ is located at the optimal injection wake phase in the second bucket of the wakefield, where $H_s-H_i$ has a maximum value. According to the trapping condition, the electrons ionized at
this wake phase are above the wake seperatrix and are trapped. During the injection process, the amplitude of the wakefield is 24 GV/m, and the beam loading effect is not observed due to the low injected electron number. After the injection pulse propagates through the krypton gas, all the 9th electrons of the krypton are ionized and trapped in the wake. The trapped number is $1.37\times10^5$ $\mu m^{-2}$, which is in good agreement with the analytical predicted trapped number $1.44\times10^5$ $\mu m^{-2}$ obtained from numerically integrating the tunneling ionization rate.
Fig.~\ref{Fig4}(d) shows the trapped electron number decreases with the decrease of the injection pulse amplitude $a_1$. PIC simulations are in good agreement with the analytical calculation. 

\begin{figure}
\begin{center}
\begin{tabular}{c}
\includegraphics[height=7cm]{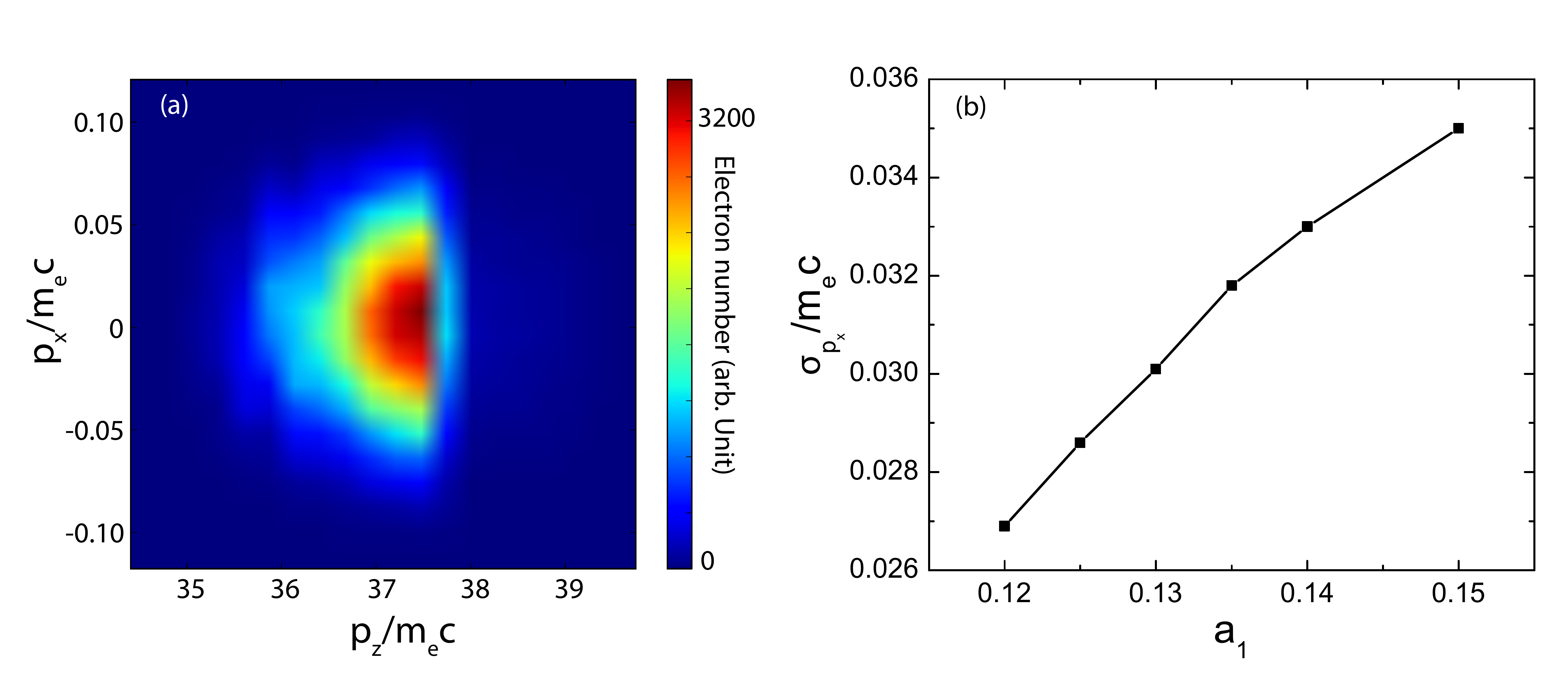}
\end{tabular}
\end{center}
\caption[example] {(Color online) (a) Transverse and longitudinal momentum distribution of the trapped electrons. (b) Root-mean-square transverse momentum versus the vector potential amplitude of the injection pulse. The electron density and the laser parameters are the same as Fig.~\ref{Fig2}. The krypton concentration is 1\%.}
\label{Fig5}
\end{figure}

Figure~\ref{Fig5}(a) shows the distribution of the trapped electrons in the transverse and longitudinal momentum space when the pump pulse propagates a distance of $1.33$ mm. The laser parameters are the same as Fig.~\ref{Fig2}. It is found that most of the electrons are ionized at the peak of the injection pulse electric field, with a zero residual transverse momentum and that some electrons are born off-peak with a finite residual transverse momentum. The maximum of the momentum is $p_x/m_e c=0.115$, which is approximately equal to the vector potential amplitude $a_1$. The rms transverse momentum of the electrons is  $\sigma_{p_x}/(m_ec)=0.035$, which is approximately equal to $0.25a_1$. This residual transverse momentum contributes to the initial transverse emittance of the electron bunch.
Since $a_1$ is an order of magnitude smaller than the vector potential amplitude used in single pulse ionization injection, the initial transverse emittance will be an order of magnitude smaller in this two-color-ionization injection scheme. Figure~\ref{Fig5}(b) shows that as the amplitude of injection pulse decreases, the rms transverse momentum of the electrons decreases; however, fewer electrons are trapped, as shown in Fig.~\ref{Fig4}(d).

\section{Summary and Discussion}\label{sec:summary}
In this paper, we have proposed a method for generation of electron bunches with low transverse momentum by using two-color laser ionization injection. 
This method can produce electron beams with an order of magnitude smaller
transverse momentum compared to single pulse ionization injection.
Simulations using a 1D PIC code were presented that show an example where the rms residual transverse momentum of the electron beam has $\sigma_{p_x}/(m_ec)\sim0.03$. In multi-dimensional geometry, not only the residual transverse momentum from the ionization process, but also the transverse momentum determined by the transverse force of the wakefield, transverse ponderomotive force of the laser and transverse bunch size will contribute to the transverse emittance. The transverse emittance can be estimated as $\epsilon_n\simeq\sigma_x\sigma_{p_x}/(m_ec)$ and
the bunch radius can be estimated as $\sigma_x\sim w_1$, where $w_1$ is the spot size of the injection pulse. By tightly focusing the injection pulse to several $\mu$m, the estimated transverse emittance can be $<0.1$ mm mrad, which is an order of magnitude smaller than that in single pulse ionization injection.

\section{Acknowledgments}
This work was supported by
the Director, Office of Science, Office of High Energy Physics, of the U.S. Department of Energy under Contract No. DE-AC02-05CH11231 and by the National Science Foundation under Grant No. PHY-0935197.

\bibliography{lule}   

\begin{thebibliography}{10}

\bibitem{Esarey09}
Esarey, E., Schroeder, C.~B., and Leemans, W.~P., ``Physics of laser-driven
  plasma-based electron accelerators,'' {\em Rev. Mod. Phys.}~{\bf 81},
  1229--1285 (2009).

\bibitem{Mangles04}
Mangles, S. P.~D., Murphy, C.~D., Najmudin, Z., Thomas, A. G.~R., Collier,
  J.~L., Dangor, A.~E., Divall, E.~J., Foster, P.~S., Gallacher, J.~G., Hooker,
  C.~J., Jaroszynski, D.~A., Langley, A.~J., Mori, W.~B., Norreys, P.~A.,
  Tsung, F.~S., Viskup, R., Walton, B.~R., and Krushelnick, K., ``Monoenergetic
  beams of relativistic electrons from intense laser-plasma interactions,''
  {\em Nature}~{\bf 431}(7008),  535--538 (2004).

\bibitem{Geddes04}
Geddes, C. G.~R., Toth, C., van Tilborg, J., Esarey, E., Schroeder, C.~B.,
  Bruhwiler, D., Nieter, C., Cary, J., and Leemans, W.~P., ``High quality
  electron beams from a plasma channel guided laser wakefield accelerator,''
  {\em Nature}~{\bf 431}(7008),  538--541 (2004).

\bibitem{Faure04}
Faure, J., Glinec, Y., Pukhov, A., Kiselev, S., Gordienko, S., Lefebvre, E.,
  Rousseau, J.-P., Burgy, F., and Malka, V., ``A laser plasma accelerator
  producing monoenergetic electron beams,'' {\em Nature}~{\bf 431}(7008),
  541--544 (2004).

\bibitem{Leemans06b}
Leemans, W.~P., Nagler, B., Gonsalves, A.~J., T\'{o}th, C., Nakamura, K.,
  Geddes, C. G.~R., Esarey, E., Schroeder, C.~B., and Hooker, S.~M., ``{GeV}
  electron beams from a centimetre-scale accelerator,'' {\em Nature Phys.}~{\bf
  2},  696--699 (2006).

\bibitem{Esarey97a}
Esarey, E., Hubbard, R.~F., Leemans, W.~P., Ting, A., and Sprangle, P.,
  ``Electron injection into plasma wake fields by colliding laser pulses,''
  {\em Phys. Rev. Lett.}~{\bf 79}(14),  2682--2685 (1997).

\bibitem{Schroeder99b}
Schroeder, C.~B., Lee, P.~B., Wurtele, J.~S., Esarey, E., and Leemans, W.~P.,
  ``Generation of ultrashort electron bunches by colliding laser pulses,'' {\em
  Phys. Rev. E}~{\bf 59}(5),  6037--6047 (1999).

\bibitem{Fubiani04}
Fubiani, G., Esarey, E., Schroeder, C.~B., and Leemans, W.~P., ``Beat wave
  injection of electrons into plasma waves using two interfering laser
  pulses,'' {\em Phys. Rev. E}~{\bf 70}(1),  016402 (2004).

\bibitem{Kotaki04}
Kotaki, H., Masuda, S., Kando, M., Koga, J.~K., and Nakajima, K., ``Head-on
  injection of a high quality electron beam by the interaction of two laser
  pulses,'' {\em Phys. Plasmas}~{\bf 11}(6),  3296--3302 (2004).

\bibitem{Faure06}
Faure, J., Rechatin, C., Norlin, A., Lifschitz, A., Glinec, Y., and Malka, V.,
  ``Controlled injection and acceleration of electrons in plasma wakefields by
  colliding laser pulses,'' {\em Nature}~{\bf 444}(7120),  737--739 (2006).

\bibitem{Rechatin09}
Rechatin, C., Faure, J., Ben-Ismail, A., Lim, J., Fitour, R., Specka, A.,
  Videau, H., Tafzi, A., Burgy, F., and Malka, V., ``Controlling the
  phase-space volume of injected electrons in a laser-plasma accelerator,''
  {\em Phys. Rev. Lett.}~{\bf 102}(16),  164801 (2009).

\bibitem{Kotaki09}
Kotaki, H., Daito, I., Kando, M., Hayashi, Y., Kawase, K., Kameshima, T.,
  Fukuda, Y., Homma, T., Ma, J., Chen, L.-M., Esirkepov, T.~Z., Pirozhkov,
  A.~S., Koga, J.~K., Faenov, A., Pikuz, T., Kiriyama, H., Okada, H.,
  Shimomura, T., Nakai, Y., Tanoue, M., Sasao, H., Wakai, D., Matsuura, H.,
  Kondo, S., Kanazawa, S., Sugiyama, A., Daido, H., and Bulanov, S.~V.,
  ``Electron optical injection with head-on and countercrossing colliding laser
  pulses,'' {\em Phys. Rev. Lett.}~{\bf 103},  194803 (2009).

\bibitem{Geddes08}
Geddes, C. G.~R., Nakamura, K., Plateau, G.~R., {Cs. T\'{o}th}, Cormier-Michel,
  E., Esarey, E., Schroeder, C.~B., Cary, J.~R., and Leemans, W.~P.,
  ``Plasma-density-gradient injection of low absolute-momentum-spread electron
  bunches,'' {\em Phys. Rev. Lett.}~{\bf 100}(21),  215004 (2008).

\bibitem{Schmid10}
Schmid, K., Buck, A., Sears, C. M.~S., Mikhailova, J.~M., Tautz, R., Herrmann,
  D., Geissler, M., Krausz, F., and Veisz, L., ``Density-transition based
  electron injector for laser driven wakefield accelerators,'' {\em Phys. Rev.
  ST Accel. Beams}~{\bf 13}(9),  091301 (2010).

\bibitem{Faure10}
Faure, J., Rechatin, C., Lundh, O., Ammoura, L., and Malka, V., ``Injection and
  acceleration of quasimonoenergetic relativistic electron beams using density
  gradients at the edges of a plasma channel,'' {\em Phys. Plasmas}~{\bf
  17}(8),  083107 (2010).

\bibitem{MinChen06}
Chen, M., Sheng, Z.-M., Ma, Y.-Y., and Zhang, J., ``Electron injection and
  trapping in a laser wakefield by field ionization to high-charge states of
  gases,'' {\em J. Appl. Phys.}~{\bf 99}(5),  056109 (2006).

\bibitem{McGuffey10}
McGuffey, C., Thomas, A. G.~R., Schumaker, W., Matsuoka, T., Chvykov, V.,
  Dollar, F.~J., Kalintchenko, G., Yanovsky, V., Maksimchuk, A., Krushelnick,
  K., Bychenkov, V.~Y., Glazyrin, I.~V., and Karpeev, A.~V., ``Ionization
  induced trapping in a laser wakefield accelerator,'' {\em Phys. Rev.
  Lett.}~{\bf 104}(2),  025004 (2010).

\bibitem{Pak10}
Pak, A., Marsh, K.~A., Martins, S.~F., Lu, W., Mori, W.~B., and Joshi, C.,
  ``Injection and trapping of tunnel-ionized electrons into laser-produced
  wakes,'' {\em Phys. Rev. Lett.}~{\bf 104}(2),  025003 (2010).

\bibitem{Liu11}
Liu, J.~S., Xia, C.~Q., Wang, W.~T., Lu, H.~Y., Wang, C., Deng, A.~H., Li,
  W.~T., Zhang, H., Liang, X.~Y., Leng, Y.~X., Lu, X.~M., Wang, C., Wang,
  J.~Z., Nakajima, K., Li, R.~X., and Xu, Z.~Z., ``All-optical cascaded laser
  wakefield accelerator using ionization-induced injection,'' {\em Phys. Rev.
  Lett.}~{\bf 107},  035001 (2011).

\bibitem{Pollock11}
Pollock, B.~B., Clayton, C.~E., Ralph, J.~E., Albert, F., Davidson, A., Divol,
  L., Filip, C., Glenzer, S.~H., Herpoldt, K., Lu, W., Marsh, K.~A., Meinecke,
  J., Mori, W.~B., Pak, A., Rensink, T.~C., Ross, J.~S., Shaw, J., Tynan,
  G.~R., Joshi, C., and Froula, D.~H., ``Demonstration of a narrow energy
  spread, ${\sim}0.5~\mathrm{GeV}$ electron beam from a two-stage laser
  wakefield accelerator,'' {\em Phys. Rev. Lett.}~{\bf 107},  045001 (2011).

\bibitem{MinChen12}
Chen, M., Esarey, E., Schroeder, C.~B., Geddes, C. G.~R., and Leemans, W.~P.,
  ``Theory of ionization-induced trapping in laser-plasma accelerators,'' {\em
  Phys. Plasmas}~{\bf 19}(3),  033101 (2012).

\bibitem{Wiggins2010}
Wiggins, S.~M., Issac, R.~C., Welsh, G.~H., Brunetti, E., Shanks, R.~P.,
  Anania, M.~P., Cipiccia, S., Manahan, G.~G., Aniculaesei, C., Ersfeld, B.,
  Islam, M.~R., Burgess, R. T.~L., Vieux, G., Gillespie, W.~A., MacLeod, A.~M.,
  van~der Geer, S.~B., de~Loos, M.~J., and Jaroszynski, D.~A., ``High quality
  electron beams from a laser wakefield accelerator,'' {\em Plasma Phys.
  Control. Fusion}~{\bf 52},  124032 (2010).

\bibitem{Plateau2012}
Plateau, G.~R., Geddes, C. G.~R., Thorn, D.~B., Chen, M., Benedetti, C.,
  Esarey, E., Gonsalves, A.~J., Matlis, N.~H., Nakamura, K., Schroeder, C.~B.,
  Shiraishi, S., Sokollik, T., van Tilborg, J., Toth, C., Trotsenko, S., Kim,
  T.~S., Battaglia, M., St\"ohlker, T., and Leemans, W.~P., ``Low-emittance
  electron bunches from a laser-plasma accelerator measured using single-shot
  x-ray spectroscopy,'' {\em Phys. Rev. Lett.}~{\bf 109},  064802 (2012).

\bibitem{Fuchs2009}
Fuchs, M., Weingartner, R., Popp, A., Major, Z., Becker, S., Osterhoff, J.,
  Cortrie, I., Zeitler, B., H\"orlein, R., Tsakiris, G.~D., Schramm, U.,
  Rowlands-Rees, T.~P., Hooker, S.~M., Habs, D., Krausz, F., Karsch, S., and
  Gr\"uner, F., ``Laser-driven soft-x-ray undulator source,'' {\em Nature
  Phys.}~{\bf 5},  826 (2009).

\bibitem{Huang2012}
Huang, Z., Ding, Y., and Schroeder, C.~B., ``Compact x-ray free-electron laser
  from a laser-plasma accelerator using a transverse-gradient undulator,'' {\em
  Phys. Rev. Lett.}~{\bf 109},  204801 (2012).

\bibitem{Maier2012}
Maier, A.~R., Meseck, A., Reiche, S., Schroeder, C.~B., Seggebrock, T., and
  Gr\"uner, F., ``Demonstration scheme for a laser-plasma-driven free-electron
  laser,'' {\em Phys. Rev. X}~{\bf 2},  031019 (2012).

\bibitem{Hidding2012}
Hidding, B., Pretzler, G., Rosenzweig, J.~B., K\"onigstein, T., Schiller, D.,
  and Bruhwiler, D.~L., ``Ultracold electron bunch generation via plasma
  photocathode emission and acceleration in a beam-driven plasma blowout,''
  {\em Phys. Rev. Lett.}~{\bf 108},  035001 (2012).

\bibitem{Popov2004}
Popov, V.~S., ``Tunnel and multiphoton ionization of atoms and ions in a strong
  laser field (keldysh theory),'' {\em Physics Uspekhi}~{\bf 47},  855--885
  (2004).

\bibitem{MinChen2013}
Chen, M., Cormier-Michel, E., Geddes, C., Bruhwiler, D.~L., Yu, L.~L., Esarey,
  E., Schroeder, C.~B., and Leemans, W.~P., ``Numerical modeling of laser
  tunneling ionization in explicit particle-in-cell codes,'' {\em J. Comput.
  Phys.}~{\bf 236},  220--228 (2013).

\bibitem{JeanLuc2012}
Vay, J.-L., Grote, D.~P., Cohen, R.~H., and Friedman, A., ``Novel methods in
  the particle-in-cell accelerator code-framework warp,'' {\em Computational
  Science and Discovery}~{\bf 5},  014019 (2012).

\end{thebibliography}
\bibliographystyle{spiebib} 

\end{document}